\documentclass{article}

\usepackage{microtype}
\usepackage{graphicx}
\usepackage{subfig}
\usepackage{amsmath}
\usepackage{booktabs} 

\usepackage{hyperref}

\usepackage[accepted]{icml2019}

\icmltitlerunning{Replica Conditional Sequential Monte Carlo}

\begin{document}

\twocolumn[
\icmltitle{Replica Conditional Sequential Monte Carlo}



\icmlsetsymbol{equal}{*}

\begin{icmlauthorlist}
\icmlauthor{Alexander Y. Shestopaloff}{ed,turing}
\icmlauthor{Arnaud Doucet}{ox,turing}
\end{icmlauthorlist}

\icmlaffiliation{turing}{The Alan Turing Institute, London, UK}
\icmlaffiliation{ox}{Department of Statistics, University of Oxford, Oxford, UK}
\icmlaffiliation{ed}{School of Mathematics, University of Edinburgh, Edinburgh, UK}

\icmlcorrespondingauthor{Alexander Y. Shestopaloff}{ashestopaloff@turing.ac.uk}

\icmlkeywords{Markov Chain Monte Carlo, State Space Models}

\vskip 0.3in
]



\printAffiliationsAndNotice{}  

\begin{abstract}
We propose a Markov chain Monte Carlo (MCMC) scheme to perform state
inference in non-linear non-Gaussian state-space models. Current state-of-the-art
methods to address this problem rely on particle MCMC techniques and
its variants, such as the iterated conditional Sequential Monte Carlo
(cSMC) scheme, which uses a Sequential Monte Carlo (SMC) type proposal
within MCMC. A deficiency of standard SMC proposals is that they only
use observations up to time $t$ to propose states at time $t$ when
an entire observation sequence is available. More sophisticated SMC
based on lookahead techniques could be used but they can be difficult
to put in practice. We propose here replica cSMC where we build SMC
proposals for one replica using information from the entire observation
sequence by conditioning on the states of the other replicas. This
approach is easily parallelizable and we demonstrate its excellent
empirical performance when compared to the standard iterated cSMC
scheme at fixed computational complexity.
\end{abstract}

\section{Introduction}

We consider discrete-time state-space models. They can be described
by a latent Markov process $(X_{t})_{t\ge1}$ and an observation process
$(Y_{t})_{t\ge1}$, $(X_{t},Y_{t})$ being $\mathcal{X}\times\mathcal{Y}$-valued,
which satisfy $X_{1}\sim\mu(\cdot)$ and
\begin{equation}
X_{t+1}|\{X_{t}=x\}\sim f(\cdot|x)\qquad Y_{t}|\{X_{t}=x\}\sim g(\cdot|x)
\end{equation}
for $t\ge1$. Our goal is to sample from posterior distribution of the latent states $X_{1:T}:=\left(X_{1},...,X_{T}\right)$
given a realization of the observations $Y_{1:T}=y_{1:T}$. This distribution admits a density given by
\begin{equation}
p(x_{1:T}|y_{1:T})\propto\mu(x_{1})g(y_{1}|x_{1})\prod_{t=2}^{T}f(x_{t}|x_{t-1})g(y_{t}|x_{t}).
\end{equation}
This sampling problem is now commonly addressed using an MCMC scheme
known as the iterated cSMC sampler \cite{Andrieu_Doucet_Holenstein_2010}
and extensions of it; see, e.g., \cite{ShestopaloffNeal2018}. This
algorithm relies on a SMC-type proposal mechanism. A limitation of
these algorithms is that they typically use data only up to time $t$
to propose candidate states at time $t$, whereas the entire sequence
$y_{1:T}$ is observed in the context we are interested in. To address
these issues, various lookahead techniques have been proposed in the
SMC literature; see \cite{Chen2013} for a review. Alternative approaches
relying on a parametric approximation of the backward information
filter used for smoothing in state-space models \cite{Briers2010}
have also been recently proposed in \cite{ScharthKohn2016,Guarniero_Lee_Johansen_2017,Ruiz_Kappen_2017,Heng2017}.
When applicable, these iterative methods have demonstrated good performance.
However, it is unclear how these ideas could be adapted to the MCMC
framework investigated here. Additionally these methods are difficult
to put in practice for multimodal posterior distributions.

In this paper, we propose a novel approach which allows us to build proposals
for cSMC that allows considering all observed data in a proposal,
based on conditioning on replicas of the state variables. Our approach
is based purely on Monte Carlo sampling, bypassing any need for approximating
functions in the estimate of the backward information filter.

The rest of this paper is organized as follows. In Section \ref{sec:Iterated-Conditional-Sequential},
we review the iterated cSMC algorithm and outline its limitations.
Section \ref{sec:Replica-Iterated-Conditional} introduces the replica
iterated cSMC methodology. In Section \ref{sec:Examples}, we demonstrate
the methodology on a linear Gaussian model, two non-Gaussian state
space models from \cite{ShestopaloffNeal2018} as well as the Lorenz-96
model from \cite{Heng2017}.

\section{Iterated cSMC\label{sec:Iterated-Conditional-Sequential}}

The iterated cSMC sampler is an MCMC method for sampling from a target
distribution of density $\pi\left(x_{1:T}\right):=\pi_T\left(x_{1:T}\right)$. It relies on a modified SMC scheme targeting a sequence of auxiliary target probability
densities $\{\pi_t\left(x_{1:t}\right)\}_{t=1,...,T-1}$ and a sequence of proposal densities $q_{1}\left(x_{1}\right)$ and $q_{t}(x_{t}|x_{t-1})$ for $t\in\{2,...,T\}$. These target densities are such that $\pi_t(x_{1:t})/\pi_{t-1}(x_{1:t-1})\propto \beta_t(x_{t-1},x_t)$.

\subsection{Algorithm}
We define the `incremental importance weights' for $t\geq2$ as
\begin{align}
w_{t}(x_{t-1},x_{t})&:=\frac{\pi_t\left(x_{1:t}\right)}{\pi_{t-1}\left(x_{1:t-1}\right)q_{t}(x_{t}|x_{t-1})}\propto\frac{\beta_t(x_{t-1},x_t)}{q_{t}(x_{t}|x_{t-1})} \label{eq:incrementalweight}
\end{align}
and for $t=1$ as
\begin{equation}
w_{1}(x_{0},x_{1}):=\frac{\pi_1(x_{1})}{q_{1}(x_{1})}.
\end{equation}
\begin{algorithm}[t]
\protect\caption{Iterated cSMC kernel $K\left(x_{1:T},x'_{1:T}\right)$~\label{alg:CSMC}}

cSMC step.
\begin{enumerate}
\item \textsf{At time} $t=1$
\begin{enumerate}
\item \textsf{Sample $b_{1}$ uniformly on $[N]$ and set} $x_{1}^{b_{1}}=x_{1}.$
\item \textsf{For }$i\in\left[N\right]\backslash\{b_{1}\}$, \textsf{sample}
$x_{1}^{i}\sim q_{1}\left(\cdot\right)$.
\item \textsf{Compute} $w_{1}(x_{0}, x_{1}^{i})$ for $i\in\left[N\right]$.
\end{enumerate}
\item \textsf{At times} $t=2,\ldots,T$
\begin{enumerate}
\item \textsf{Sample $b_{t}$ uniformly on $[N]$ and set} $x_{t}^{b_{t}}=x_{t}$.
\item \textsf{For }$i\in\left[N\right]\backslash\{b_{t}\}$, \textsf{sample
}\\$a_{t-1}^{i}\sim$ Cat$\{ w_{t-1}(x_{t-2}^{a_{t-2}^{j}},x_{t-1}^{j});j\in[N]\}$.
\item \textsf{For }$i\in\left[N\right]\backslash\{b_{t}\}$, \textsf{sample
}$x_{t}^{i}\sim q_{t}(\left.\cdot\right\vert x_{t-1}^{a_{t-1}^{i}})$\textsf{.}
\item \textsf{Compute} $w_{t}(x_{t-1}^{a_{t-1}^{i}}, x_{t}^{i})$ for $i\in\left[N\right]$.
\end{enumerate}
\end{enumerate}
Backward sampling step.
\begin{enumerate}
\item \textsf{At times} $t=T$
\begin{enumerate}
\item \textsf{Sample }$b_{T}\sim$ Cat$\{w_{T}(x_{T-1}^{a_{T-1}^{j}},x_{T}^{j});j\in[N]\}$.
\end{enumerate}
\item \textsf{At times} $t=T-1,...,1$
\begin{enumerate}
\item \textsf{Sample }$b_{t}\sim$ \\ Cat$\{\beta_{t+1}(x_t^j, x_{t+1}^{b_{t+1}})w_{t}(x_{t-1}^{a_{t-1}^{j}},x_{t}^{j});j\in[N]\}$.
\end{enumerate}
\end{enumerate}
Output $x'_{1:T}=x_{1:T}^{b_{1:T}}:=\left(x_{1}^{b_{1}},\ldots,x_{T}^{b_{T}}\right)$.
\end{algorithm}
We introduce a dummy variable $x_{0}$ to simplify notation.
We let $N\geq2$ be the number of particles used by the algorithm and $[N]:=\{1,...,N\}$.
We introduce the notation $\mathbf{x}_{t}=\left(x_{t}^{1},\ldots,x_{t}^{N}\right)\in\mathcal{X}^{N},$
$\mathbf{a}_{t}=\left(a_{t}^{1},\ldots,a_{t}^{N}\right)\in\left\{ 1,\ldots,N\right\} ^{N}$,
$\mathbf{x}_{1:T}=(\mathbf{x}_{1},\mathbf{x}_{2},...,\mathbf{x}_{T}),$
$\mathbf{a}_{1:T-1}=(\mathbf{a}_{1},\mathbf{a}_{2},...,\mathbf{a}_{T-1}$)
and $\mathbf{x}_{t}^{-b_{t}}=\mathbf{x}_{t}\backslash x_{t}^{b_{t}}$,
$\mathbf{x}_{1:T}^{-b_{1:T}}=\left\{ \mathbf{x}_{1}^{-b_{1}},\ldots,\mathbf{x}_{T}^{-b_{T}}\right\} $,
$\mathbf{a}_{t-1}^{-b_{t}}=\mathbf{a}_{t-1}\backslash a_{t-1}^{b_{t}}$,
$\mathbf{a}_{1:T-1}^{-b_{2:T}}=\left\{ \mathbf{a}_{1}^{-b_{2}},\ldots,\mathbf{a}_{T-1}^{-b_{T}}\right\} $
and set $b_{t}=a_{t}^{b_{t+1}}$ for $t=1,...,T-1.$ \\
It can be shown that the iterated cSMC kernel, described in Algorithm \ref{alg:CSMC},
is invariant w.r.t. $\pi(x_{1:T})$. Given the current state
$x_{1:T}$, the cSMC step introduced in \cite{Andrieu_Doucet_Holenstein_2010}
samples from the following distribution
\begin{align}
\Phi(\left.\mathbf{x}_{1:T}^{-b_{1:T}},\mathbf{a}_{1:T-1}^{-b_{2:T}}\right\vert x_{1:T}^{b_{1:T}},b_{1:T})=\delta_{x_{1:T}}\left(x_{1:T}^{b_{1:T}}\right)\notag \\ \times {\displaystyle \prod\limits _{i=1,i\neq b_{1}}^{N}}q_1\left(x_{1}^{i}\right)\,{\displaystyle \prod\limits _{t=2}^{T}}\thinspace{\displaystyle \prod\limits _{i=1,i\neq b_{t}}^{N}}\lambda(\left.a_{t-1}^{i},x_{t}^{i}\right\vert \mathbf{x}_{t-1}),\label{eq:CPF}
\end{align}
where
\begin{align}
\lambda\left(\left.a_{t-1}^{i}=k,x_{t}^{i}\right\vert \mathbf{x}_{t-1}\right)&=\frac{w_{t-1}(x_{t-2}^{a_{t-1}^{k}},x_{t-1}^{k})}{\sum_{j=1}^{N}w_{t-1}(x_{t-2}^{a_{t-1}^{j}},x_{t-1}^{j})}~ \notag \\ &\times q_{t}(\left.x_{t}^{i}\right\vert x_{t-1}^{k}).
\end{align}
This can be combined to a backward sampling step introduced in \cite{Whiteley2010}; see \cite{Finke2016,ShestopaloffNeal2018} for a detailed derivation. It can be shown that the combination of these two steps defined a Markov kernel that preserves the following extended target distribution
\begin{align}
\gamma(\mathbf{x}_{1:T},\mathbf{a}_{1:T-1}^{-b_{2:T}},b_{1:T}):=\frac{\pi(x_{1:T}^{b_{1:T}})}{N^{T}} \notag \\ \times \ \Phi(\left.\mathbf{x}_{1:T}^{-b_{1:T}},\mathbf{a}_{1:T-1}^{-b_{2:T}}\right\vert x_{1:T}^{b_{1:T}},b_{1:T})\label{eq:extended}
\end{align}
as invariant distribution. In particular, it follows that if $x_{1:T}\sim \pi$ then $x'_{1:T}\sim \pi$. The algorithm is described in Algorithm 1 where we use the notation Cat$\{c_i;i\in[N]\}$ to denote the categorical distribution of probabilities $p_i\propto c_i$.

Iterated cSMC has been widely adopted for state space models, i.e. when the target is $\pi(x_{1:T})=p(x_{1:T}|y_{1:T})$. The default sequence of auxiliary targets one uses is $\pi_t(x_{1:t})=p(x_{1:t}|y_{1:t})$ for $t=1,...,T-1$ resulting in the incremental importance weights
\begin{equation}
w_{t}(x_{t-1},x_{t})\propto\frac{f(x_{t}|x_{t-1})g(y_{t}|x_{t})}{q_{t}(x_{t}|x_{t-1})}\label{eq:incrementalweight-2}
\end{equation}
for $t \geq 2$ and
\begin{equation}
w_{1}(x_{0},x_{1})\propto\frac{\mu(x_{1})g(y_{1}|x_{1})}{q_{1}(x_{1})}
\end{equation}
for $t=1$. Typically we will attempt to select a proposal which minimizes the
variance of the incremental weight, which at time $t \geq 2$ is $q_{t}^{\mathrm{opt}}(x_{t}|x_{t-1})=p(x_{t}|x_{t-1},y_{t})\propto g(y_{t}|x_{t})f(x_{t}|x_{t-1})$
or an approximation of it.

\subsection{Limitations of Iterated cSMC}

When using the default sequence of auxiliary targets for state space models, iterated cSMC does not exploit
a key feature of the problem at hand. The cSMC step typically uses
a proposal at time $t$ that only relies on the observation $y_{t}$,
i.e. $q_{t}(x_{t}|x_{t-1})=p\left(x_{t}|x_{t-1},y_{t}\right)$, as
it targets at time $t$ the posterior density $p\left(x_{1:t}|y_{1:t}\right)$.
In high-dimensions and/or in the presence of highly informative observations,
the discrepancy between successive posterior densities $\{p\left(x_{1:t}|y_{1:t}\right)\}_{t\geq1}$
will be high. Consequently the resulting importance weights $\{w_{t}(x_{t-1}^{a_{t-1}^{i}},x_{t}^{i});i\in[N]\}$
will have high variance and the resulting procedure will be inefficient.

Ideally one would like to use the sequence of marginal
smoothing densities as auxiliary densities, that is $\pi_t(x_{1:t})=p\left(x_{1:t}|y_{1:T}\right)$ for $t=1,...,T-1$.
Unfortunately, this is not possible as $p\left(x_{1:t}|y_{1:T}\right)\propto p\left(x_{1:t}|y_{1:t-1}\right)p\left(y_{t:T}|x_{t}\right)$
cannot be evaluated pointwise up to a normalizing constant. To address
this problem in a standard SMC framework, recent contributions \cite{ScharthKohn2016,Guarniero_Lee_Johansen_2017,Ruiz_Kappen_2017,Heng2017}
perform an analytical approximation $\hat{p}\left(y_{t:T}|x_{t}\right)$
of the backward information filter $p\left(y_{t:T}|x_{t}\right)$
based on an iterative particle mechanism and target instead $\{\hat{p}\left(x_{1:t}|y_{1:T}\right)\}_{t\geq1}$
where $\hat{p}\left(x_{1:t}|y_{1:T}\right)\propto p\left(x_{1:t}|y_{1:t-1}\right)\hat{p}\left(y_{t:T}|x_{t}\right)$
using proposals of the form $q_{t}\left(x_{t}|x_{t-1}\right)\propto f\left(x_{t}|x_{t-1}\right)\hat{p}\left(y_{t:T}|x_{t}\right)$.
These methods can perform well but it requires a careful design of
the analytical approximation and is difficult to put in practice for
multimodal posteriors. Additionally, it is unclear how these could
be adapted in an iterated cSMC framework without introducing any bias.

Versions of iterated cSMC using an independent approximation to the
backward information filter based on Particle Efficient Importance
Sampling \cite{ScharthKohn2016} have been proposed \cite{GrotheKleppeLiesenfeld}
though they still require a choice of analytical approximation and
use an approximation to the backward information filter which is global. This can become inefficient in high dimensional state scenarios.

\section{Replica Iterated cSMC\label{sec:Replica-Iterated-Conditional}}
We introduce a way to directly use the iterated cSMC algorithm to
target a sequence of approximations $\{\hat{p}\left(x_{1:t}|y_{1:T}\right)\}_{t\geq1}$ to the marginal smoothing densities of a state space
model. Our proposed method is based on sampling from a target over
multiple copies of the space as done in, for instance, the Parallel
Tempering or Ensemble MCMC \cite{Neal2011} approaches. However, unlike
in these techniques, we use copies of the space to define a sequence
of intermediate distributions in the cSMC step informed by the whole
dataset. This enables us to draw samples of $X_{1:T}$ that incorporate
information about all of the observed data.  Related recent work includes
\cite{Leimkuhler2018}, where information sharing amongst an ensemble of
replicas is used to improve MCMC proposals.

\subsection{Algorithm}
We start by defining the replica target for some $K\geq2$ by
\begin{align}
\bar{\pi}(x_{1:T}^{(1:K)})=\prod_{k=1}^{K}p(x_{1:T}^{(k)}|y_{1:T}).
\end{align}
Each of the replicas $x_{1:T}^{(k)}$ is updated in turn by running
Algorithm \ref{alg:CSMC} with a different sequence of intermediate
targets which we describe here. Consider updating replica $k$ and let $\hat{p}^{(k)}(y_{t+1:T}|x_{t})$
be an estimator of the backward information filter,
built using replicas other than the $k$-th one, $x_{t+1}^{(-k)}=(x_{t+1}^{(1)},\ldots,x_{t+1}^{(k-1)},x_{t+1}^{(k+1)},\ldots,x_{t+1}^{(K)}).$
For convenience of notation, we take $\hat{p}^{(k)}(y_{T+1:T}|x_{T}):=1$. At time $t$, the cSMC does
target approximation of the marginal smoothing distribution $p\left(x_{1:t}|y_{1:T}\right)$ as in \cite{ScharthKohn2016,Guarniero_Lee_Johansen_2017,Ruiz_Kappen_2017,Heng2017}.
This is of the form $\hat{p}^{(k)}\left(x_{1:t}|y_{1:T}\right)\propto p\left(x_{1:t}|y_{1:t}\right)\hat{p}^{(k)}\left(y_{t+1:T}|x_{t}\right)$.
This means that the cSMC for replica $k$ uses the novel incremental weights at time $t\geq2$
\begin{align}
w_{t}^{\left(k\right)}(x_{t-1},x_{t}) &:= \frac{\hat{p}^{(k)}\left(x_{1:t}|y_{1:T}\right)}{\hat{p}^{(k)}\left(x_{1:t-1}|y_{1:T}\right)q_{t}(x_{t}|x_{t-1})}  \\&\propto\frac{g(y_{t}|x_{t})f(x_{t}|x_{t-1})\hat{p}^{(k)}\left(y_{t+1:T}|x_{t}\right)}{\hat{p}^{(k)}\left(y_{t:T}|x_{t-1}\right)q_{t}(x_{t}|x_{t-1})} \notag 
\end{align}
and $w_{1}^{\left(k\right)}(x_{0},x_{1})\propto g(y_{1}|x_{1})\mu(x_{1})\hat{p}^{(k)}(y_{t+1:T}|x_{1})/q_{1}(x_{1})$. We would like to use the proposal minimizing the variance of the incremental
weight, which at time $t\geq2$ is $q_{t}^{\mathrm{opt}}(x_{t}|x_{t-1})\propto g(y_{t}|x_{t})f(x_{t}|x_{t-1})\hat{p}^{(k)}\left(y_{t+1:T}|x_{t}\right)$ or an approximation of it.

The full replica cSMC update for $\bar{\pi}$ is described in Algorithm
\ref{alg:replica-CSMC} and is simply an application of Algorithm \ref{alg:CSMC} to
a sequence of target densities for each replica. A proof of the validity of the algorithm is provided
in the Supplementary Material.
\begin{algorithm}
\protect\caption{Replica cSMC update~\label{alg:replica-CSMC}}

For $k=1,\ldots,K$
\begin{enumerate}
\item \textsf{Build an approximation $\hat{p}^{(k)}\left(y_{t+1:T}|x_{t}\right)$ of $p\left(y_{t+1:T}|x_{t}\right)$ using the replicas $(x_{t+1}^{(1)'},\ldots,x_{t+1}^{(k-1)'},x_{t+1}^{(k+1)},\ldots,x_{t+1}^{(K)})$ \hspace{-0.2cm}for $t=1,...,T-1$}.
\item \textsf{Run Algorithm \ref{alg:CSMC} with target $\pi(x_{1:T}) = p(x_{1:T}|y_{1:T})$ and auxiliary targets
$\pi_{t}(x_{1:t}) = \hat{p}^{(k)}\left(x_{1:t}|y_{1:T}\right)$ for $t = 1,\ldots, T-1$ with initial state $x_{1:T}^{(k)}$ to return $x_{1:T}^{(k')}$}.
\end{enumerate}
Output $x_{1:T}^{(1:K)'}$.
\end{algorithm}

One sensible way to initialize the replicas is to set them to sequences sampled from standard independent
SMC passes. This will start the Markov chain not too far from equilibrium. For multimodal
distributions, initialization is particularly crucial, since we need to ensure that different
replicas are well-distributed amongst the various modes at the start of the run.

\subsection{Setup and Tuning\label{subsec:Setup-and-Tuning}}

The replica cSMC sampler requires an estimator $\hat{p}^{(k)}\left(y_{t+1:T}|x_{t}\right)$ of
the backward information filter based on $x_{t+1}^{(-k)}$. For our algorithm, we propose an
estimator $\hat{p}^{(k)}\left(y_{t+1:T}|x_{t}\right)$ that is not
based on any analytical approximation of $p\left(y_{t+1:T}|x_{t}\right)$
but simply on a Monte Carlo approximation built using the other replicas,
\begin{equation}
\hat{p}^{(k)}\left(y_{t+1:T}|x_{t}\right)\propto\sum_{j\neq k}\frac{f(x_{t+1}^{\left(j\right)}|x_{t})}{p(x_{t+1}^{\left(j\right)}|y_{1:t})},\label{eq:approxMonteCarlobackward}
\end{equation}
where $p\left(x_{t+1}|y_{1:t}\right)$ denotes the predictive density
of $x_{t+1}$. The rationale for this approach is that at equilibrium
the components of $x_{t+1}^{(-k)}$ are an iid sample from a product
of $K-1$ copies of the smoothing density, $p\left(x_{t+1}|y_{1:T}\right)$.
Therefore, as $K$ increases, (\ref{eq:approxMonteCarlobackward})
converges to
\begin{align}
&\int\frac{f\left(x_{t+1}|x_{t}\right)}{p\left(x_{t+1}|y_{1:t}\right)}p\left(x_{t+1}|y_{1:T}\right)dx_{t+1} \notag \\
& \propto\int f\left(x_{t+1}|x_{t}\right)p\left(y_{t+1:T}|x_{t+1}\right)dx_{t+1} \notag \\
& =p\left(y_{t+1:T}|x_{t}\right). \label{eq:backwardFilter}
\end{align}
In practice, the predictive density is also unknown and we need to use an approximation
of it. Whatever being the approximation $\hat{p}\left(x_{t+1}|y_{1:t}\right)$
of $p\left(x_{t+1}|y_{1:t}\right)$ we use, the algorithm is valid. We note that for
$K = 2$, any approximation of the predictive density results in the same
incremental importance weights.

We propose to approximate the predictive density in (\ref{eq:backwardFilter}) by a constant over the entire latent space, i.e. $\hat{p}(x_{t+1}|y_{1:t}) = 1$. We justify this choice as follows. If we assume that we have informative observations, which is typical in many state space modelling scenarios, then $p(x_{t+1}|y_{1:T})$ will tend to be much more concentrated than $p(x_{t+1}|y_{1:t})$. Thus, over the region where the posterior has high density, the predictive density will be approximately constant relative to the posterior density. This suggests approximating the predictive density in (\ref{eq:backwardFilter}) by its mean with respect to the posterior density,
\begin{align}
&\int\frac{f\left(x_{t+1}|x_{t}\right)}{p\left(x_{t+1}|y_{1:t}\right)}p\left(x_{t+1}|y_{1:T}\right)dx_{t+1} \notag \\
&\approx \frac{\int f\left(x_{t+1}|x_{t}\right)p\left(x_{t+1}|y_{1:T}\right)dx_{t+1}}{\int p\left(x_{t+1}|y_{1:t}\right)p\left(x_{t+1}|y_{1:T}\right)dx_{t+1}} \notag \\
&\approx \frac{\frac{1}{K}\sum_{k=1}^{K} f(x_{t+1}^{(k)}|x_{t})}{\frac{1}{K}\sum_{k=1}^{K} p(x_{t+1}^{(k)}|y_{1:t})}. \label{eq:ConstPred}
\end{align}
Since the importance weights in cSMC at each time are defined up to a constant, sampling is not affected by the specific value of $\frac{1}{K}\sum_{k=1}^{K} p(x_{t+1}^{(k)}|y_{1:t})$. Therefore, when doing computation it can simply be set to any value, which is what we do.

We note that while the asymptotic argument doesn't hold for the estimator
in (\ref{eq:ConstPred}), when the variance of the predictive density
is greater than the variance of the posterior density, we expect the estimators
in (\ref{eq:approxMonteCarlobackward}) and (\ref{eq:ConstPred})
to be close for any finite $K$.

An additional benefit to approximating the predictive density by a constant is reduction in the variance of the mixture weights in (\ref{eq:approxMonteCarlobackward}). To see why this can be the case, consider
the following example. Suppose the predictive density of $x_{t+1}$ is $\mathcal{N}(\mu,\sigma_{0}^{2})$ and the posterior density is $\mathcal{N}(0,\sigma_{1}^{2})$, where $\sigma_{1}^{2} < \sigma_{0}^{2}$. Computing the variance of the mixture weight, we get
\begin{align}
&\textnormal{Var}\bigg(\frac{1}{p(x_{t+1}|y_{1:t})}\biggr) \notag \\
& = \frac{2\pi\sigma_{0}^{2}}{\sqrt{2\sigma_{1}^{2}\nu_{1}}}\exp\biggr[\mu^{2}\biggl(\frac{1}{\sigma_{0}^{2}}+\frac{1}{(\sigma_{0}^{2})^{2}\nu_{1}}\biggr)\biggr] \notag \\
& - \frac{2\pi\sigma_{0}^{2}}{\sigma_{1}^{2}\nu_{2}}\exp\biggr[\mu^{2}\biggl(\frac{1}{\sigma_{0}^{2}}+\frac{1}{(\sigma_{0}^{2})^{2}\nu_{2}}\biggr)\biggr].
\end{align}
where
\begin{equation}
\nu_{1}=\biggl(\frac{1}{2\sigma_{1}^{2}}-\frac{1}{\sigma_{0}^{2}}\biggr) \qquad
\nu_{2}=\biggl(\frac{1}{\sigma_{1}^{2}}-\frac{1}{\sigma_{0}^{2}}\biggr).
\end{equation}

From this we can see
that variance increases exponentially with the squared difference of predictive
and posterior means, $\mu^{2}$. As a result, we can get outliers in the mixture
weight distribution. If this happens, many of the replicas will end
up having low weights in the mixture. This will reduce the effective
number of replicas used. Using a constant approximation will weight
all of the replicas uniformly, and allow us to construct better proposals,
as illustrated in Section \ref{subsec:A-Linear-Gaussian}.

A natural extension of the proposed method is to update some of the
replicas with other than replica cSMC updates. Samples from these
replicas can then be used in estimates of the backward information
filter when doing a replica cSMC update. This makes it possible to
parallelize the method, at least to some extent. For instance, one
possibility is to do parallel independent cSMC updates on some of
the replicas.

Performing other than replica cSMC updates on some of the replicas
can be useful in multimodal scenarios. If all replicas are located
in an isolated mode, and the replica cSMC updates use an estimate
of the backward information filter based on replicas in that mode,
then the overall Markov chain will tend not to transition well to
other modes. Using samples from other types of updates in the estimate
of the backward information filter can help counteract this effect
by making transitions to other high-density regions possible.

\section{Examples\label{sec:Examples}}

We consider four models to illustrate the performance of our method.
In all examples, we assume that the model parameters are known. The
first is a simple linear Gaussian model. We use this model to demonstrate that it is sensible to use a constant
approximation to the predictive density in our estimator of the backward information
filter. We also use the linear Gaussian model to better understand the accuracy and performance of
replica cSMC. The second model, from \cite{ShestopaloffNeal2018}, demonstrates
that our proposed replica cSMC method is competitive with existing state-of-the-art methods at drawing latent state sequences
in a unimodal context. The third model, also from \cite{ShestopaloffNeal2018},
demonstrates that by updating some replica coordinates with a standard
iterated cSMC kernel, our method is able to efficiently handle multimodal
sampling without the use of specialized ``flip'' updates. The fourth
model is the Lorenz-96 model from \cite{Heng2017}, which has very
low observation noise, making it a challenging case for standard iterated cSMC.

To do our computations, we used MATLAB on an OS X system, running on
an Intel Core i5 1.3 GHz CPU. As a performance metric for the sampler,
we used autocorrelation time, which is a measure of approximately
how many steps of an MCMC chain are required to obtain the equivalent
of one independent sample. The autocorrelation time is estimated based
on a set of runs as follows. First, we estimate the overall mean using
all of the runs. Then, we use this overall mean to estimate autocovariances
for each of the runs. The autocovariance estimates are then averaged
and used to estimate the autocorrelations $\hat{\rho}_{k}$. The autocorrelation
time is then estimated as $1+2\sum_{m=1}^{M}\hat{\rho}_{m}$ where
$M$ is chosen such that for $m>M$ the autocorrelations are approximately
$0$. Code to reproduce the experiments is provided \href{https://github.com/ayshestopaloff/replicacsmc}{here}.

\subsection{A Linear Gaussian Model\label{subsec:A-Linear-Gaussian}}

Let $X_{t}=(X_{1,t}, \ldots,X_{d,t})'$ for $t=1, \ldots, T$. The latent process for this model is defined as $X_{1} \sim \mathcal{N}(0,\Sigma_{1})$, $X_{t}|\{X_{t-1}=x_{t-1}\} \sim \mathcal{N}(\Phi x_{t-1},\Sigma)$ for $t=2,\ldots,T$, where
\begin{eqnarray*}
\Phi & = &
\setlength{\arraycolsep}{1pt}
\begin{pmatrix}\phi_{1} & 0 & \cdots & 0\\
0 & \phi_{2} & \ddots & \vdots\\
\vdots & \ddots & \phi_{d-1} & 0\\
0 & \cdots & 0 & \phi_{d}
\end{pmatrix},
\quad \Sigma  =
\setlength{\arraycolsep}{1pt}
\begin{pmatrix}1 & \rho & \cdots & \rho\\
\rho & 1 & \ddots & \vdots\\
\vdots & \ddots & 1 & \rho\\
\rho & \cdots & \rho & 1
\end{pmatrix},\\
\Sigma_{1} & = &
\setlength{\arraycolsep}{1pt}
\begin{pmatrix}\sigma^{2}_{1,1} & \rho \sigma_{1,1} \sigma_{1,2}& \cdots & \rho \sigma_{1,1}\sigma_{1,d}\\
\rho \sigma_{1,2} \sigma_{1,1} & \sigma^{2}_{1,2} & \ddots & \vdots\\
\vdots & \ddots & \sigma^{2}_{1,d-1} & \rho \sigma_{1,d-1} \sigma_{1,d}\\ \rho \sigma_{1,d} \sigma_{1,1} & \cdots & \rho \sigma_{1,d} \sigma_{1,d-1} & \sigma^{2}_{1,d}
\end{pmatrix},
\end{eqnarray*}
with $\sigma^{2}_{1,i} = 1/(1-\phi_{i}^{2})$ for $i=1,\ldots,d$. The observations are $Y_{i,t}|\{X_{i,t}=x_{i,t}\} \sim \mathcal{N}(x_{i,t},1)$ for  $i=1,\ldots,d$ and $t=1,\ldots,T$. We set $T=250,d=5$ and the model's parameters to $\rho=0.7$ and
$\phi_{i}=0.9$ for $i=1,\ldots,d$. We generate a sequence from this
model to use for our experiments.


Since this is a linear Gaussian model, we are able to compute the
predictive density in (\ref{eq:approxMonteCarlobackward}) exactly
using a Kalman filter. So for replica $k$, we can use the following importance densities,
\begin{align}
q_{1}(x_{1}) & \propto\mu(x_{1})\sum_{j\neq k}\frac{f(x_{2}^{(j)}\vert x_{1})}{p(x_{2}^{(j)}|y_{1})},\nonumber \\
q_{t}(x_{t}\vert x_{t-1}) & \propto f(x_{t}\vert x_{t-1})\sum_{j\neq k}\frac{f(x_{t+1}^{(j)}\vert x_{t})}{p(x_{t+1}^{(j)}|y_{1:t})},\nonumber \\
q_{T}(x_{T}|x_{T-1}) & \propto f(x_{T}\vert x_{T-1}),\label{eq:ImportanceDensities}
\end{align}
where $t=2,\ldots,T-1$. Since these densities are Gaussian mixtures,
they can be sampled from exactly. However, as pointed out in the previous section, this approach can 
be inefficient. We will show experimentally that using a constant
approximation to the predictive density in (\ref{eq:approxMonteCarlobackward})
actually improves performance.
In all experiments, we intialize
all replicas to a sample from an independent SMC pass with the same
number of particles as used for cSMC updates. Also, the different runs in
our experiments use different random number generator seeds. 

We first check that our replica method produces answers that agree
with the posterior mean computed by a Kalman smoother. To do this,
we do $10$ replica cSMC runs with $100$ particles and $2$ replicas
for $25,000$ iterations, updating each replica conditional on the
other. We then look at whether the posterior mean of $x_{i,t}$ computed
using a Kalman smoother lies within two standard errors of the overall
mean of $10$ replica cSMC runs. We find this happens for about $91.4\%$
of the $x_{i,t}$. This indicates strong agreement between the answers
obtained by replica cSMC and the Kalman smoother.

Next, we investigate the effect of using more replicas. To do this,
we compare replica cSMC using $2$ versus $75$ replicas. We do $5$
runs of each sampler. Both samplers use $100$ particles and we do
a total of $5,000$ iterations per run. For the sampler using $75$
replicas, we update replica $1$ at every iteration and replicas $2$
to $75$ in sequence at every $20$-th iteration. For the sampler
using $2$ replicas, we update both replicas at every iteration. In
both samplers, we update replica $1$ with replica cSMC and the remaining
replica(s) with iterated cSMC. After discarding 10\% of each run as
burn-in, we use all runs for a sampler to compute autocorrelation
time.

\begin{figure}[t]
\centering
\subfloat[Replica cSMC, $2$ replicas.\label{fig:Replica-2}]
{\begin{centering}
\includegraphics[width=0.23\textwidth]{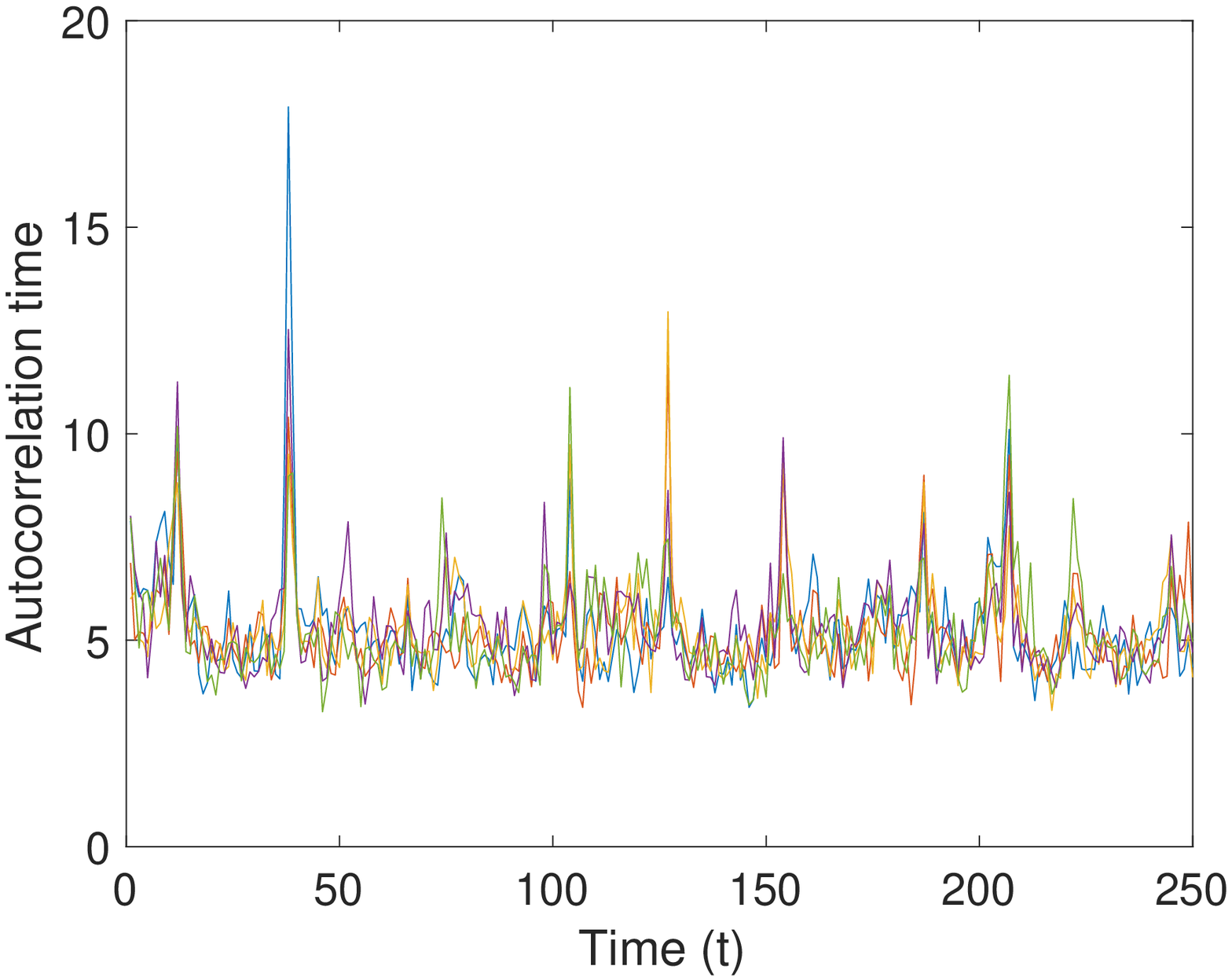}
\par\end{centering}}
\subfloat[Replica cSMC, $75$ replicas.\label{fig:Replica-75}]
{\begin{centering}
\includegraphics[width=0.23\textwidth]{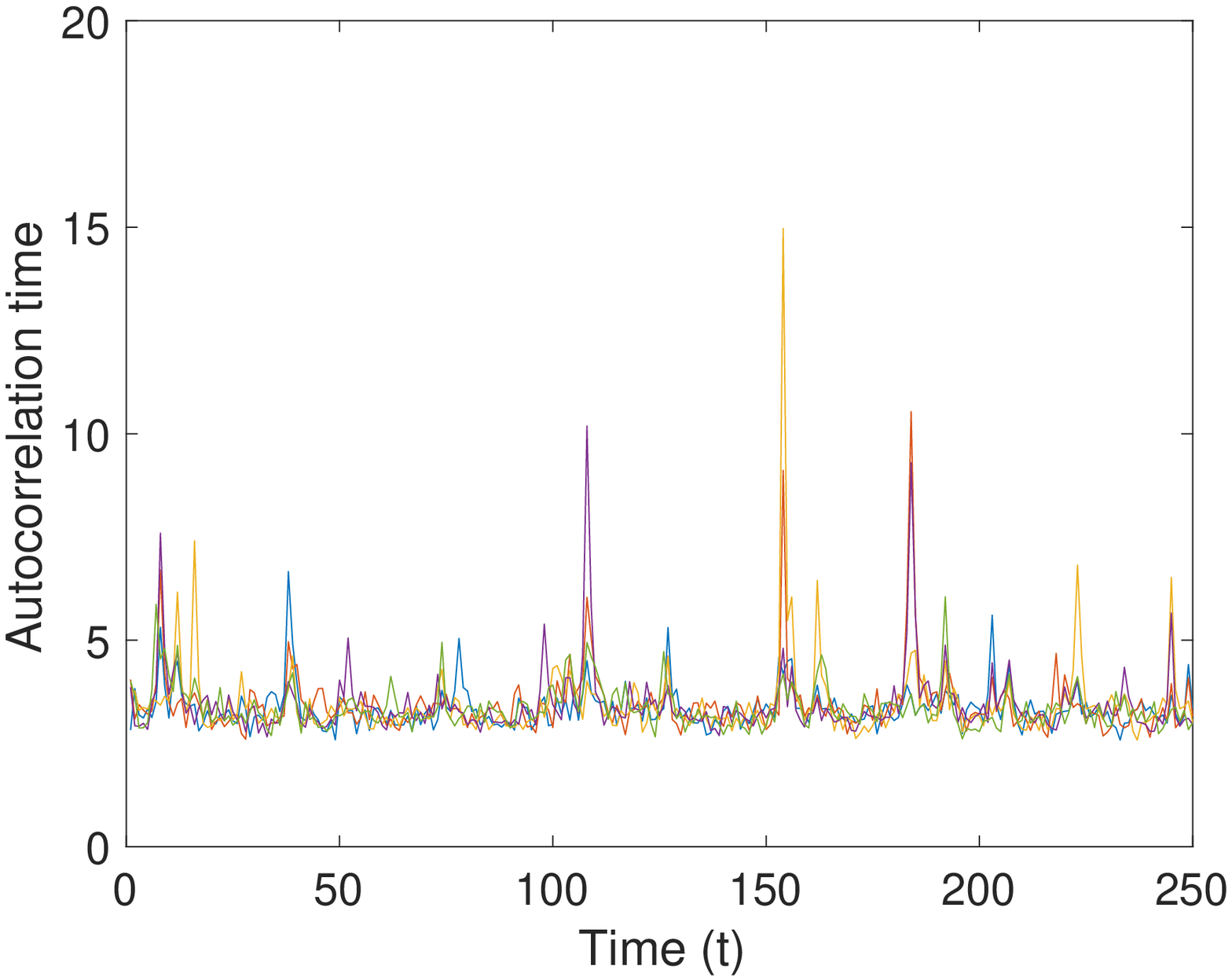}
\par\end{centering}}\\
\subfloat[Replica cSMC, $75$ replicas, constant approximation to predictive.\label{fig:Replica-approx}]
{\begin{centering}
\includegraphics[width=0.23\textwidth]{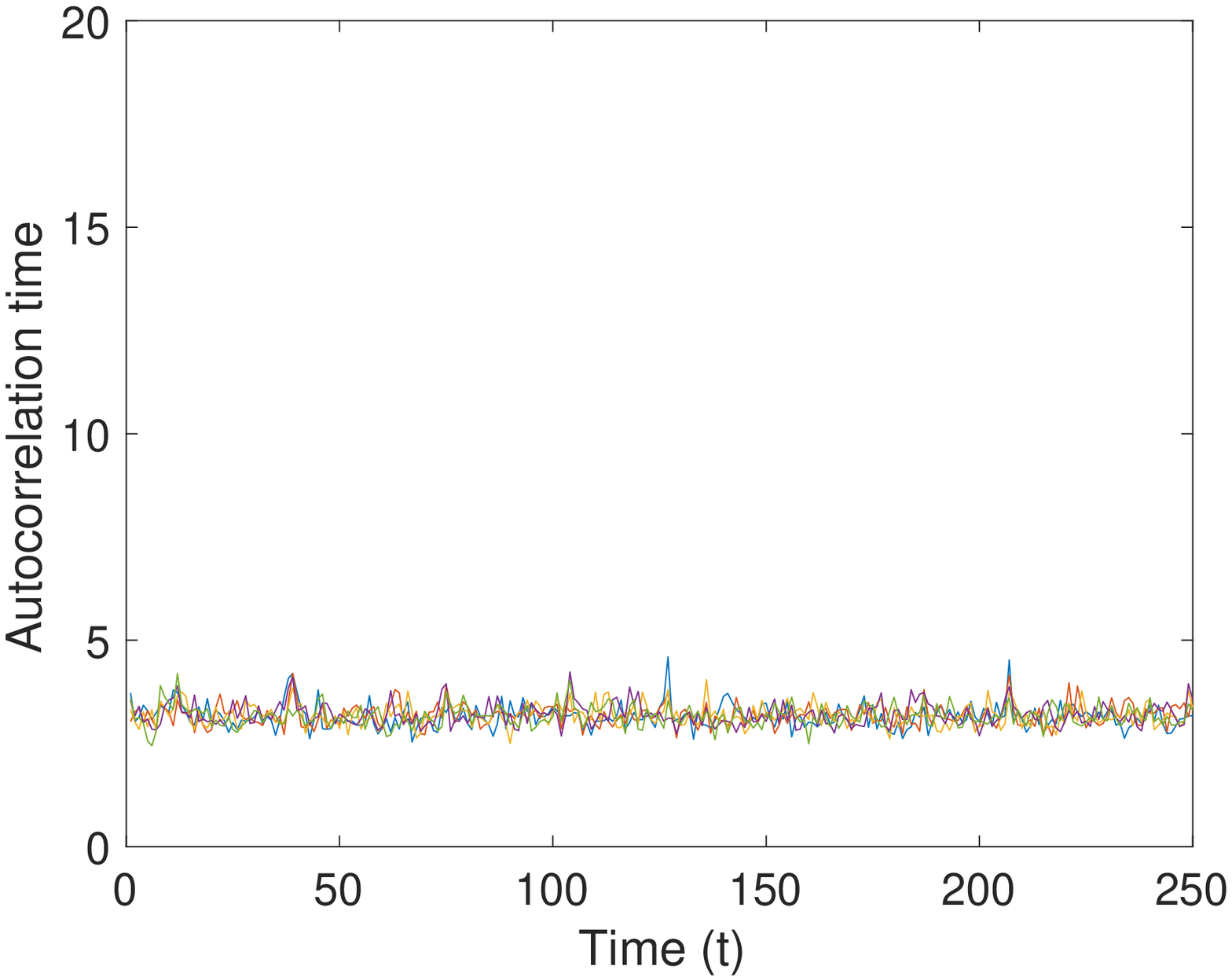}
\par\end{centering}}
\caption{Estimated autocorrelation times for each latent variable. Different
coloured lines correspond to different latent state components. The
$x$-axis corresponds to different times.\label{fig:Estimated-autocorrelation-times}}
\end{figure}

We can clearly see in Figures \ref{fig:Replica-2} and \ref{fig:Replica-75}
that using more replicas improves performance, before adjusting for
computation time. We note that for this simple example, there is no
benefit from using replica cSMC with a large number of replicas if
we take into account computation time.

To check the performance of using the constant approximation versus
the exact predictive density, we run replica cSMC with $75$ replicas
and the same settings as earlier, except using a constant approximation
to the predictive density. Figure \ref{fig:Replica-approx} shows
that using a constant approximation to the predictive density results
in peformance better than when using the true predictive density.
This is consistent with our discussion in Section \ref{subsec:Setup-and-Tuning}.

The linear Gaussian model can also be used to demonstrate that due
to looking ahead, a fixed level of precision can be achieved with
much fewer particles with replica cSMC than with standard iterated
cSMC. In scenarios where the state is high dimensional and the observations
are informative, it is difficult to efficiently sample the variables
$x_{i,1}$ with standard iterated cSMC using the initial density as
the proposal. We do $20$ runs of $2,500$ iterations of both iterated
cSMC with $700$ particles and of replica cSMC with $35$ particles
and $2$ replicas, with each replica updated given the other. We then
use the runs to estimate the standard error of the overall mean over
$20$ runs. For the variable $x_{1,1}$ sampled with iterated cSMC
we estimate the standard error to be approximately $0.0111$ whereas
for replica cSMC the estimated standard error is a similar $0.0081$,
achieved using only 5\% of the particles.

\begin{figure}
\begin{centering}
\includegraphics[width=0.23\textwidth]{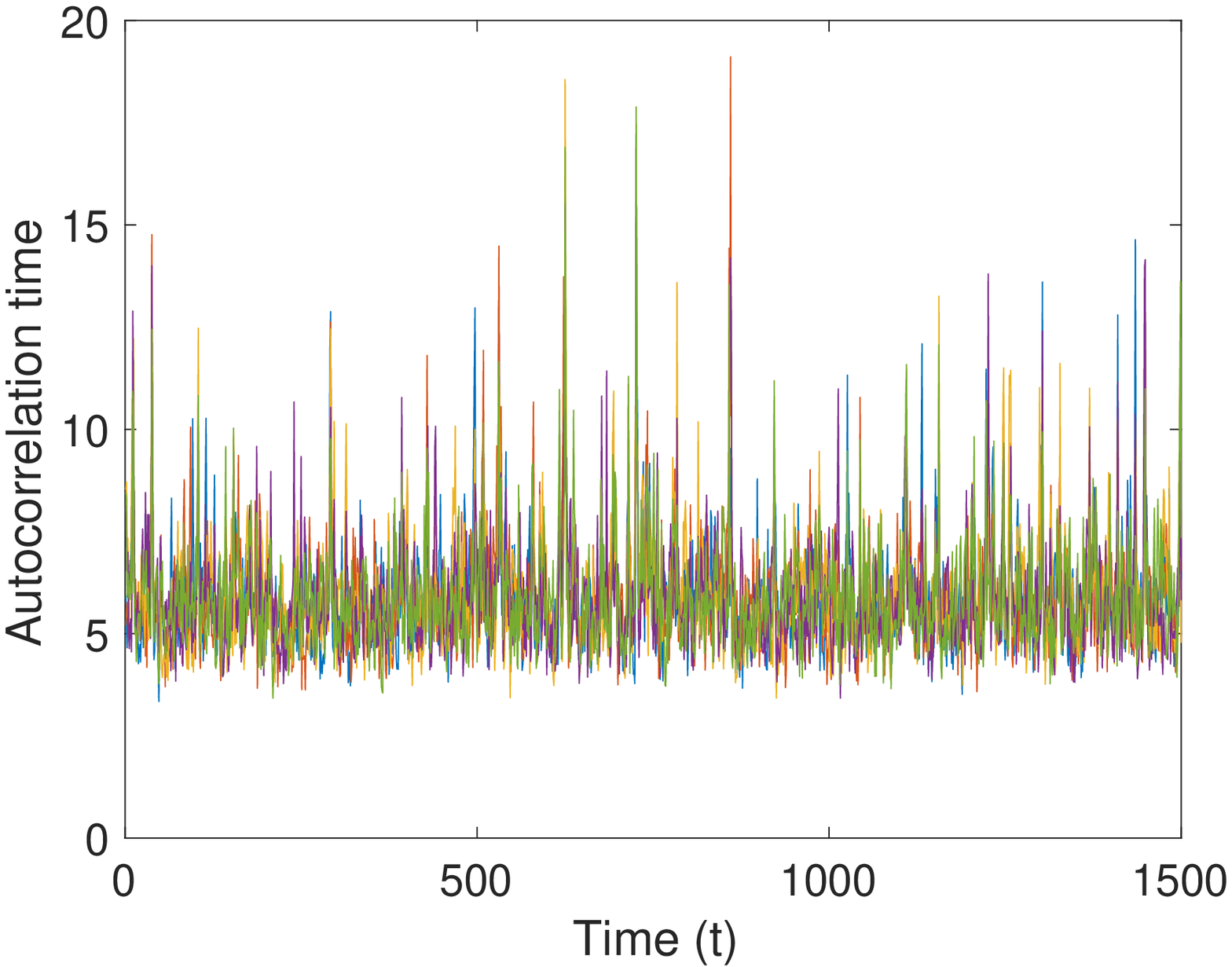}
\par\end{centering}
\caption{Estimated autocorrelation times for each latent variable. Different
coloured lines correspond to different latent state components. The
$x$-axis corresponds to different times.\label{fig:Estimated-autocorrelation-times-1}}
\end{figure}

Finally, we verify that the proposed method works well on longer time
series by running it on the linear Gaussian model but with the length
of the observed sequence set to $T=1,500$. We use $2$ replicas,
each updated given the other, and do $5$ runs of $5,000$ iterations
of the sampler to estimate the autocorrelation time for sampling the
latent variables. In Figure \ref{fig:Estimated-autocorrelation-times-1}
we can see that the replica cSMC method does not suffer from a decrease
in performance when used on longer time series.

\subsection{Two Poisson-Gaussian Models}

In this example, we consider the two models from \cite{ShestopaloffNeal2018}.
Model $1$ uses the same latent process as Section \ref{subsec:A-Linear-Gaussian}
with $T=250$, $d=10$ and $Y_{i,t}|\{X_{i,t}=x_{i,t}\} \sim \mathcal{\textnormal{Poisson}}(\exp(c+\sigma x_{i,t}))$ for $i=1,\ldots,d$ and $t=1,\ldots,T$ where $c=-0.4$ and $\sigma=0.6$. For Model $2$, we again use the
latent process in Section \ref{subsec:A-Linear-Gaussian}, with $T=500,d=15$
and $Y_{i,t}|\{X_{i,t}=x_{i,t}\} \sim \mathcal{\textnormal{Poisson}}(\sigma|x_{i,t}|))$ for $i=1,\ldots,d$ and $t=1,\ldots,T$ where $\sigma=0.8$. We assume the observations are independent given the latent states.

\begin{figure}
\centering
\subfloat[Data for Model 1.]{\centering{}\includegraphics[width=0.23\textwidth]{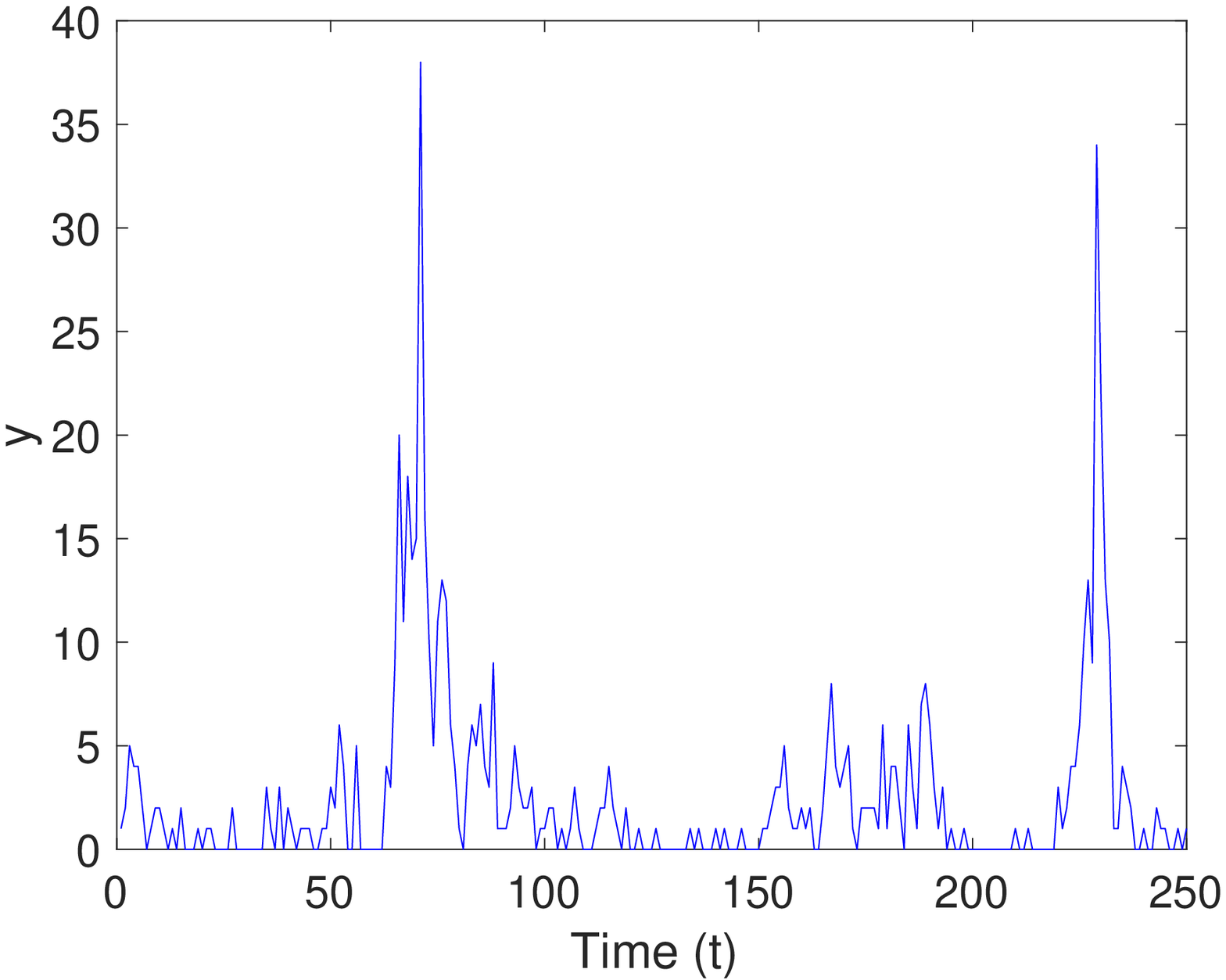}}
\subfloat[Data for Model 2.]{\centering{}\includegraphics[width=0.23\textwidth]{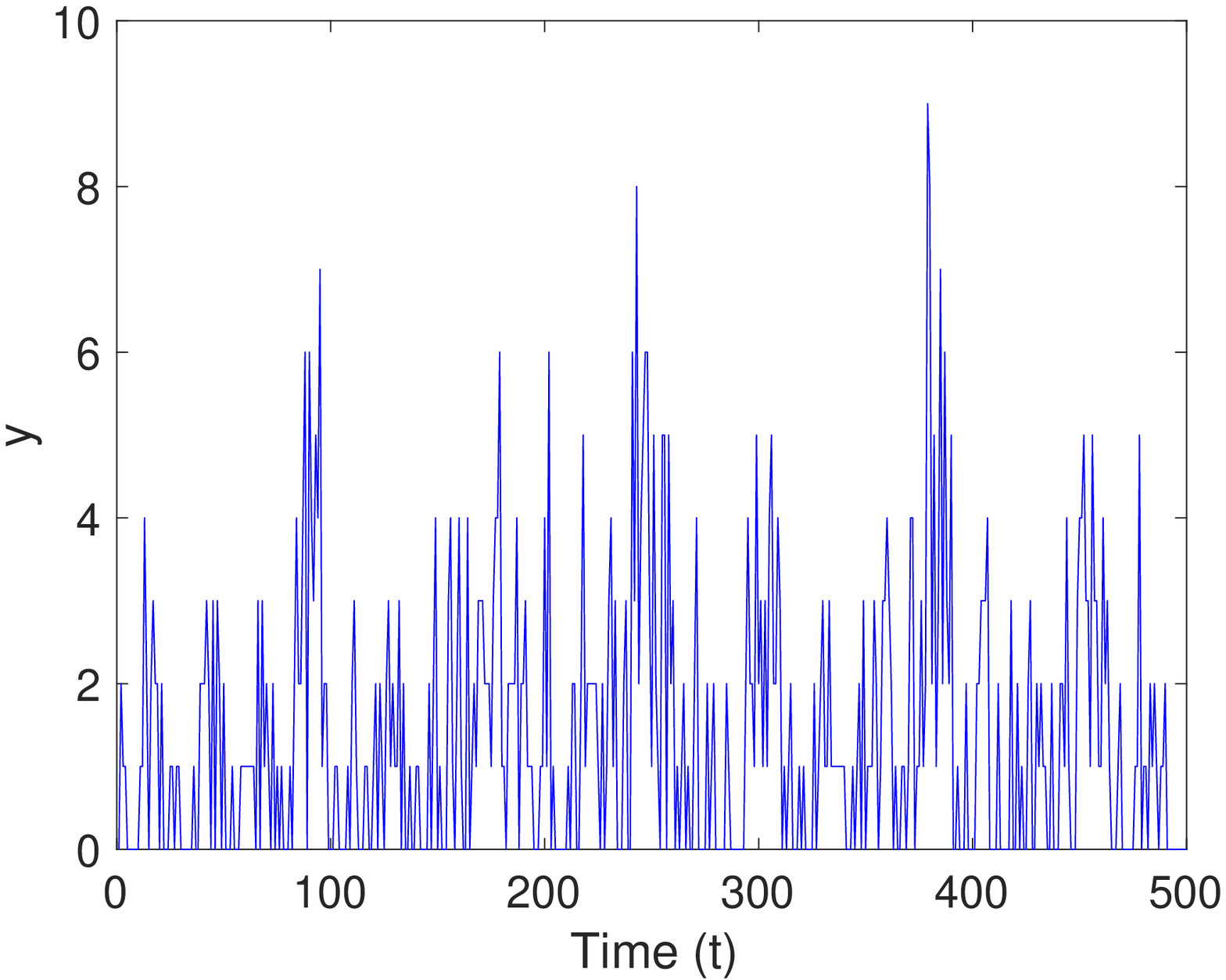}}
\caption{Simulated data from the Poisson-Gaussian models.\label{fig:Simulated-data-from}}
\end{figure}

We generate one sequence of observations from
each model. A plot of the simulated data along dimension $i=1$ is shown in Figure \ref{fig:Simulated-data-from}.
We set the importance densities $q_{t}$ for the replica cSMC sampler
to the same ones as in Section \ref{subsec:A-Linear-Gaussian}, with a
constant approximation to the predictive density.

\subsubsection*{Model 1}

We use replica cSMC with $5$ replicas, updating one replica conditional
on the other. We start with both sequences initialized to $\mathbf{0}.$
We set the number of particles to $200$. We do a total of $5$ runs
of the sampler with $5,000$ iterations, each run with a different
random number generator seed. Each iteration of replica cSMC takes approximately $0.80$ seconds.
We discard 10\% of each run as burn-in.

\begin{figure}
\subfloat[Iterated cSMC+Metropolis.]{\begin{centering}
\includegraphics[width=0.23\textwidth]{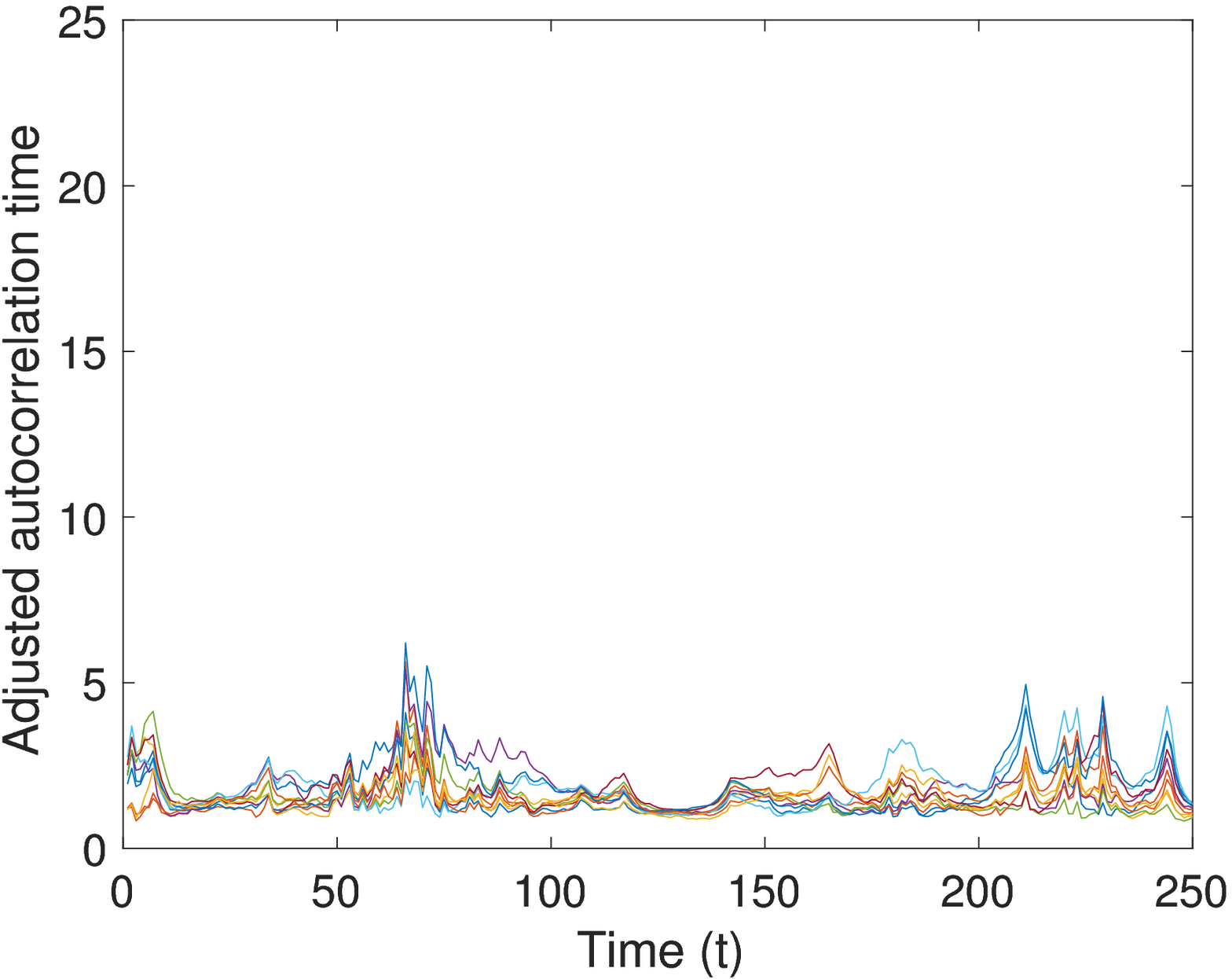}
\par\end{centering}
}\subfloat[Replica cSMC.]{\begin{centering}
\includegraphics[width=0.23\textwidth]{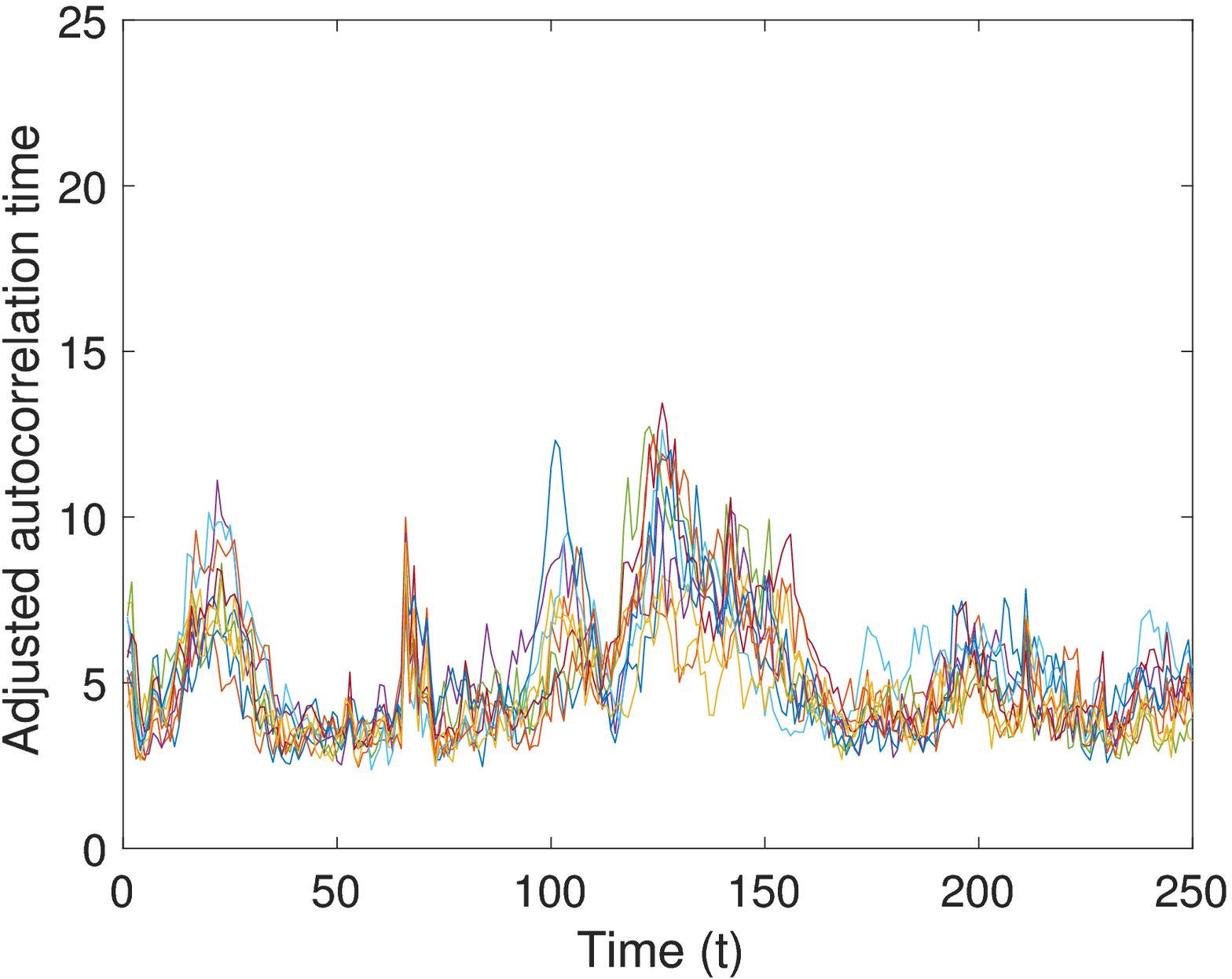}
\par\end{centering}}
\centering{}\caption{Model 1. Estimated autocorrelation times for each latent variable,
adjusted for computation time. Different coloured lines corresponds
to different latent state components. The $x$-axis corresponds to
different times.\label{fig:Model1acf}}
\end{figure}

Plots of autocorrelation time comparing replica cSMC to the best method
in \cite{ShestopaloffNeal2018} for sampling each of the latent variables
are shown in Figure \ref{fig:Model1acf}. The benchmark method takes approximately
$0.21$ seconds per iteration. We can see that the proposed
replica cSMC method performs relatively well when compared to their
best method after adjusting for computation time. The figure for iterated
cSMC+Metropolis was reproduced using code available with \cite{ShestopaloffNeal2018}.

\subsubsection*{Model 2}
For this model, the challenge is to move between the many different
modes of the latent state due to conditioning on $|x_{i,t}|$ in the
observation density. The marginal posterior of $x_{i,t}$ has two
modes and is symmetric around $0$. Additional modes appear due to
uncertainty in the signs of state components.

We use a total of $50$ replicas and update $49$ of the $50$ replicas
with iterated cSMC and one replica with replica cSMC. This is done
to prevent the Markov chain from being stuck in a single mode while
at the same time enabling the replica cSMC update to use an estimate
of the backward information filter based on replicas that are distributed
across the state space. We initialize all replicas using sequences
drawn from independent SMC passes with $1,000$ particles, and run
the sampler for a total of $2,000$ iterations. Both replica cSMC
and iterated cSMC updates use $100$ particles.

\begin{figure}
\centering
\subfloat[Trace plot for $x_{1,300}$.]{\centering{}\includegraphics[width=0.23\textwidth]{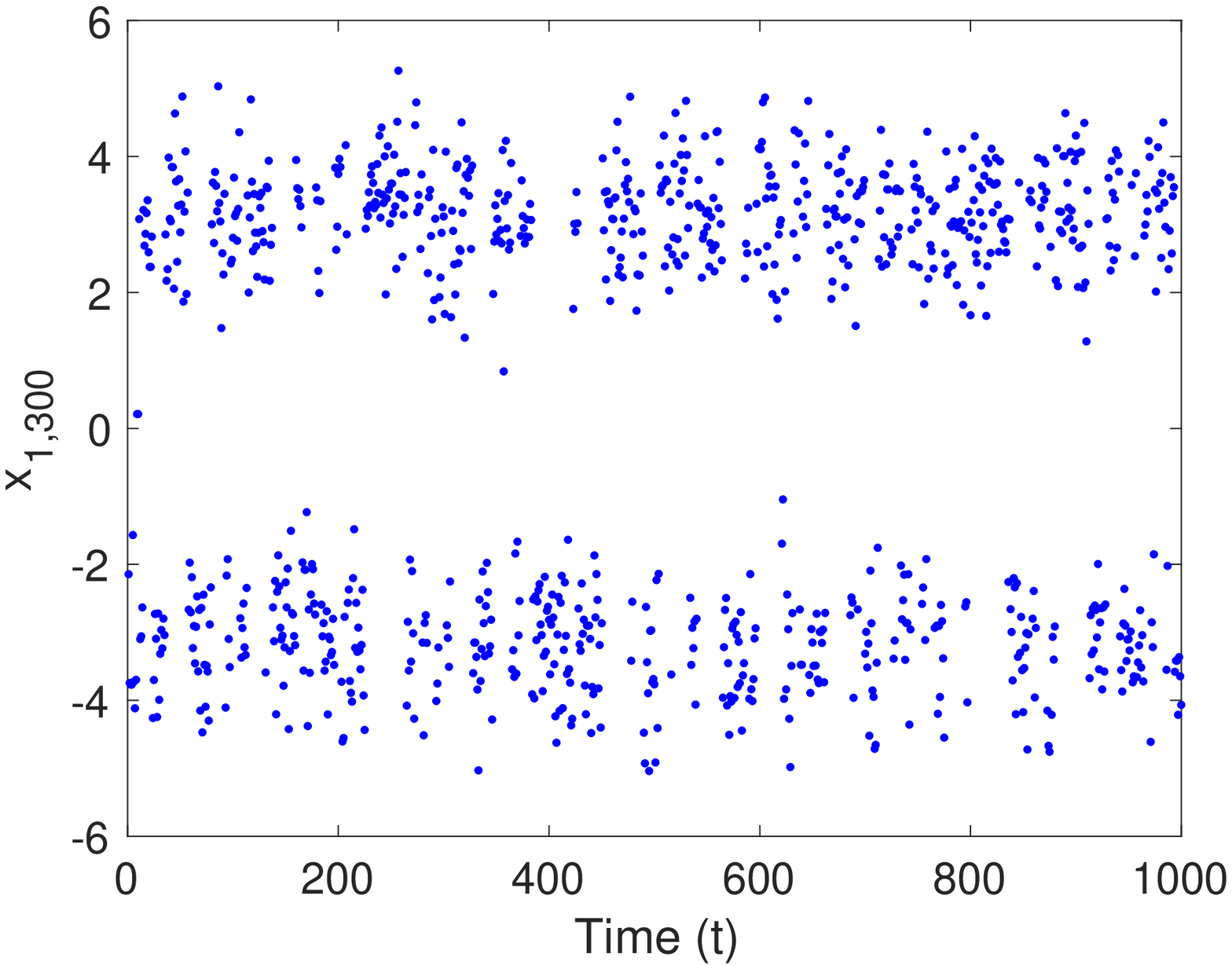}}
\subfloat[Trace plot for $x_{3,208}x_{4,208}$.]{\centering{}\includegraphics[width=0.23\textwidth]{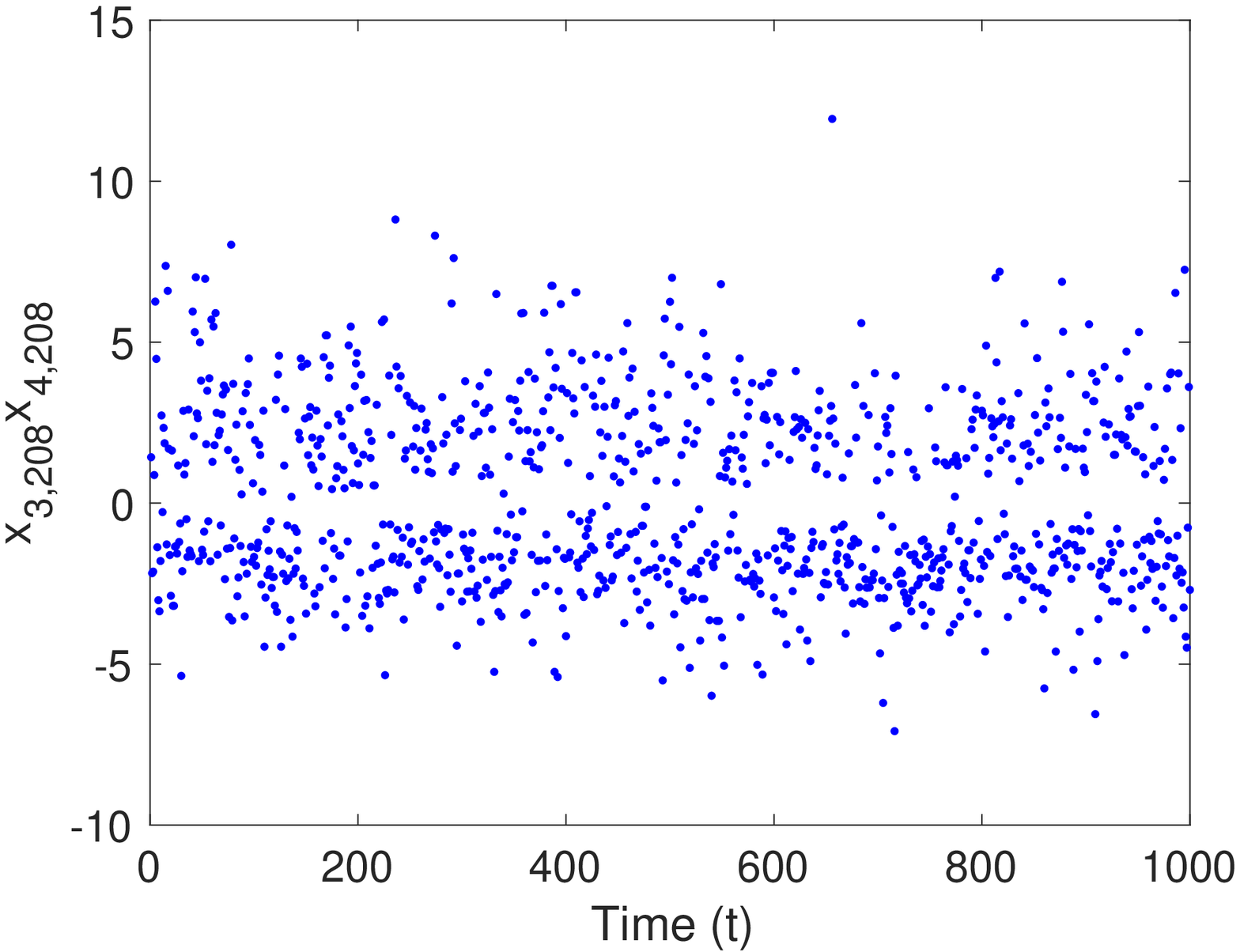}}
\caption{Trace plots for Model 2.\label{fig:Model2trace}}
\end{figure}

In Figure \ref{fig:Model2trace} we plot every other sample of the
same functions of state as in \cite{ShestopaloffNeal2018} of the
replica updated with replica cSMC. This is the the coordinate $x_{1,300}$
with true value $-1.99$ and $x_{3,208}x_{4,208}$ with true value
$-4.45$. The first has two well-separated modes and the second is
ambiguous with respect to sign. We see that the sampler is able to
explore different modes, without requiring any specialized ``flip''
updates or having to use a much larger number of particles, as is
the case in \cite{ShestopaloffNeal2018}.

We note that the replicas doing iterated cSMC updates tend to get
stuck in separate modes for long periods of time, as expected. However,
as long as these replicas are well-distributed across the state space
and eventually explore it, the bias in the estimate of the backward
information filter will be low and vanish asymptotically. The samples
from the replica cSMC update will consequently be a good approximation
to samples from the target density. Further improvement of the estimate
of the backward information filter based on replicas in multimodal
scenarios remains an open problem.

\subsection{Lorenz-96 Model}

Finally, we look at the Lorenz-96 model in a low-noise regime from
\cite{Heng2017}. The state function for this model is the It\^{o} process
$\xi(s)=(\xi_{1}(s),\ldots,\xi_{d}(s))$ defined as the weak solution
of the stochastic differential equation (SDE)
\begin{equation}
\textnormal{d}\xi_{i}=(-\xi_{i-1}\xi_{i-2}+\xi_{i-1}\xi_{i+1}-\xi_{i}+\alpha)\textnormal{d}t+\sigma_{f}\textnormal{d}B_{i}\label{eq:Lorenz}
\end{equation}
for $i=1,\ldots,d$, where indices are computed modulo $d$, $\alpha$
is a forcing parameter, $\sigma_{f}^{2}$ is a noise parameter and
$B(s)=(B_{1}(s),\ldots,B_{d}(s))$ is $d$-dimensional standard Brownian
motion. The initial condition for the SDE is $\xi(0)=\mathcal{N}(\mathbf{0},\sigma_{f}^{2}\mathcal{I}_{d})$.
We observe the process on a regular grid of size $h>0$ as $Y_{t}\sim\mathcal{N}(H\xi(th),R)$,
where $t=0,\ldots,T$. We will assume that the process is only partially
observed, with $H_{ii}=1$ for $i=1,\ldots,p$ and $0$ otherwise,
for $p=d-2$.

\begin{figure}
\begin{centering}
\includegraphics[width=0.23\textwidth]{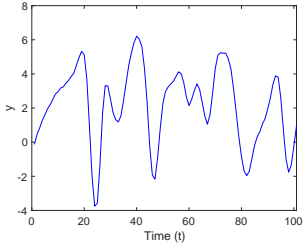}
\par\end{centering}
\caption{Simulated data from Lorenz-96 model along coordinate $i=1$.\label{fig:Simulated-data-from-Lorenz-96}}
\end{figure}

We discretize the SDE (\ref{eq:Lorenz}) by numerically integrating
the drift using a fourth-order Runge-Kutta scheme and adding Brownian
increments. Let $u$ be the mapping obtained by numerically integrating
the drift of (\ref{eq:Lorenz}) on $[0,h]$. This discretization produces
a state space model with $X_{1}\sim\mathcal{N}(\mathbf{0},\sigma_{f}^{2}\mathcal{I})$, $X_{t}|\{X_{t-1}=x_{t-1}\}\sim\mathcal{N}(u(x_{t-1}),\sigma_{f}^{2}h\mathcal{I})$ for $t=2,\ldots,T+1$ and $Y_{t}|\{X_{t}=x_{t}\}\sim\mathcal{N}(Hx_{t},R)$
for $t=1,\ldots,T+1$. We set $d=16,\sigma_{f}^{2}=10^{-2},R=10^{-3}\mathcal{I}_{p}$
and $\alpha=4.8801$. The process is observed for $10$ time units,
which corresponds to $h=0.1$, $T=100$, and a step size of $10^{-2}$
for the Runge-Kutta scheme. A plot of data generated from the Lorenz-96
model along one of the coordinates is shown in Figure \ref{fig:Simulated-data-from-Lorenz-96}.

We compare the performance of replica cSMC with two replicas, updating
each replica conditional on the other, to an iterated cSMC scheme.
For iterated cSMC, we use the model's initial density as $q_{1}$
and the model's transition density as $q_{t}$ for $t\geq2$. For
replica cSMC, we use the following importance densities for replica
$k$,
\begin{align}
q_{1}(x_{1}) & \propto f(x_{1})\sum_{j\neq k}\phi(x_{1}|x_{2}^{(j)}),\nonumber \\
q_{t}(x_{t}|x_{t-1}) & \propto f(x_{t}|x_{t-1})\sum_{j\neq k}\phi(x_{t}|x_{t+1}^{(j)}),\nonumber \\
q_{T}(x_{T}|x_{T-1}) & \propto f(x_{T}|x_{T-1}),\label{eq:Case1}
\end{align}
where $t=2,\ldots,T-1$ and $\phi$ is the $p$-dimensional Gaussian density with mean $Hu^{-1}(x_{t+1}^{(j)})$ and variance
$\sigma_{f}^{2}h\mathcal{I}_{p}$, that is, the mean is computed by
running the Runge-Kutta scheme backward in time starting at the replica
state $x_{t+1}^{(j)}$. We initialize the iterated cSMC sampler and
each replica in the replica cSMC sampler with a sequence drawn from
an independent SMC pass with $3,000$ particles. We run replica cSMC
with $200$ particles for $30,000$ iterations ($0.7$ seconds per
iteration) and compare to standard iterated cSMC with $600$ particles,
which we also run for $30,000$ iterations ($0.7$ seconds per iteration),
thus making the computational time equal.

\begin{figure}
\begin{centering}
\subfloat[Standard cSMC trace, $x_{1,45}$.]{\begin{centering}
\includegraphics[width=0.23\textwidth]{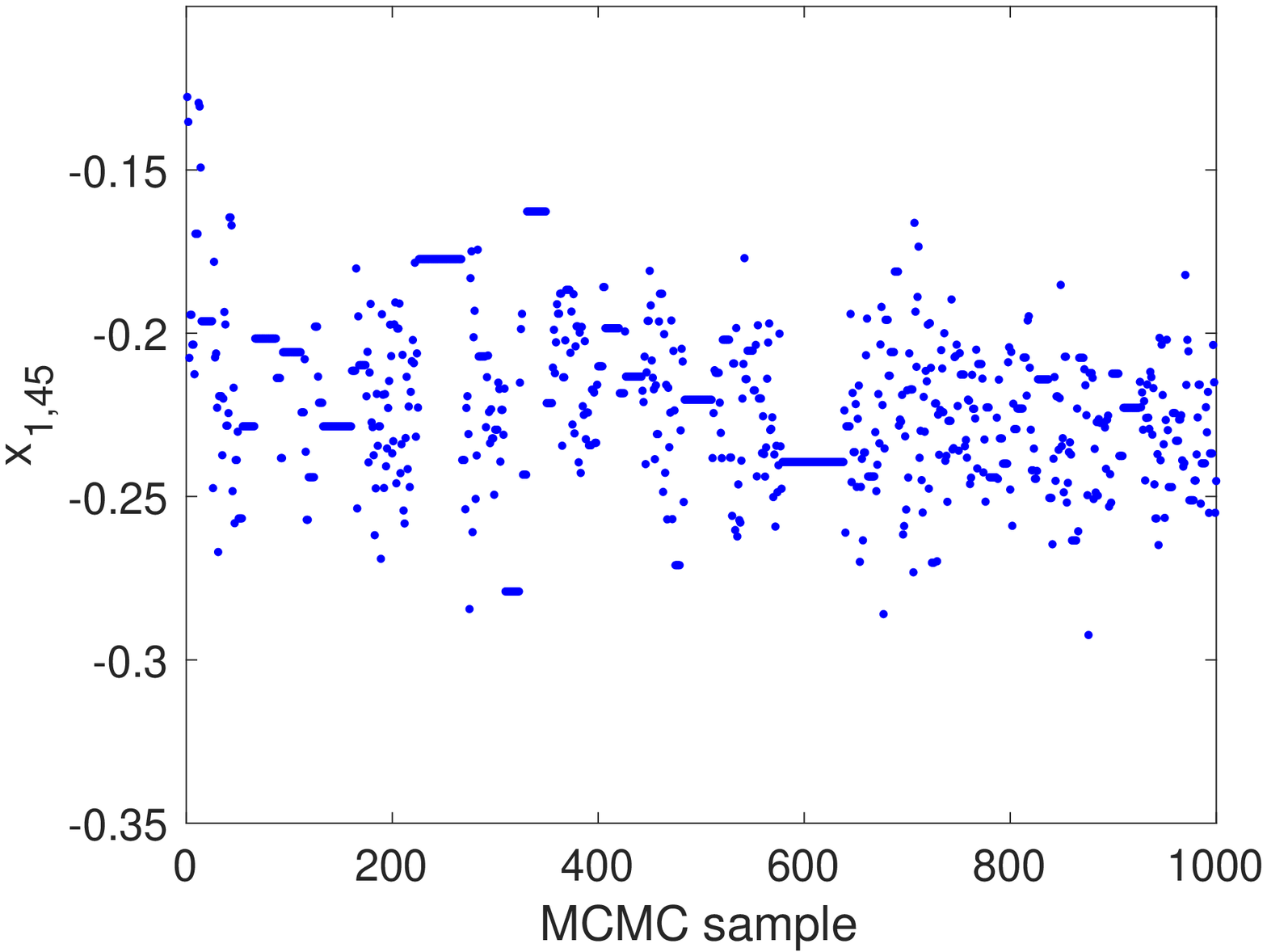}
\par\end{centering}
}\subfloat[Replica cSMC trace, $x_{1,45}$.]{\begin{centering}
\includegraphics[width=0.23\textwidth]{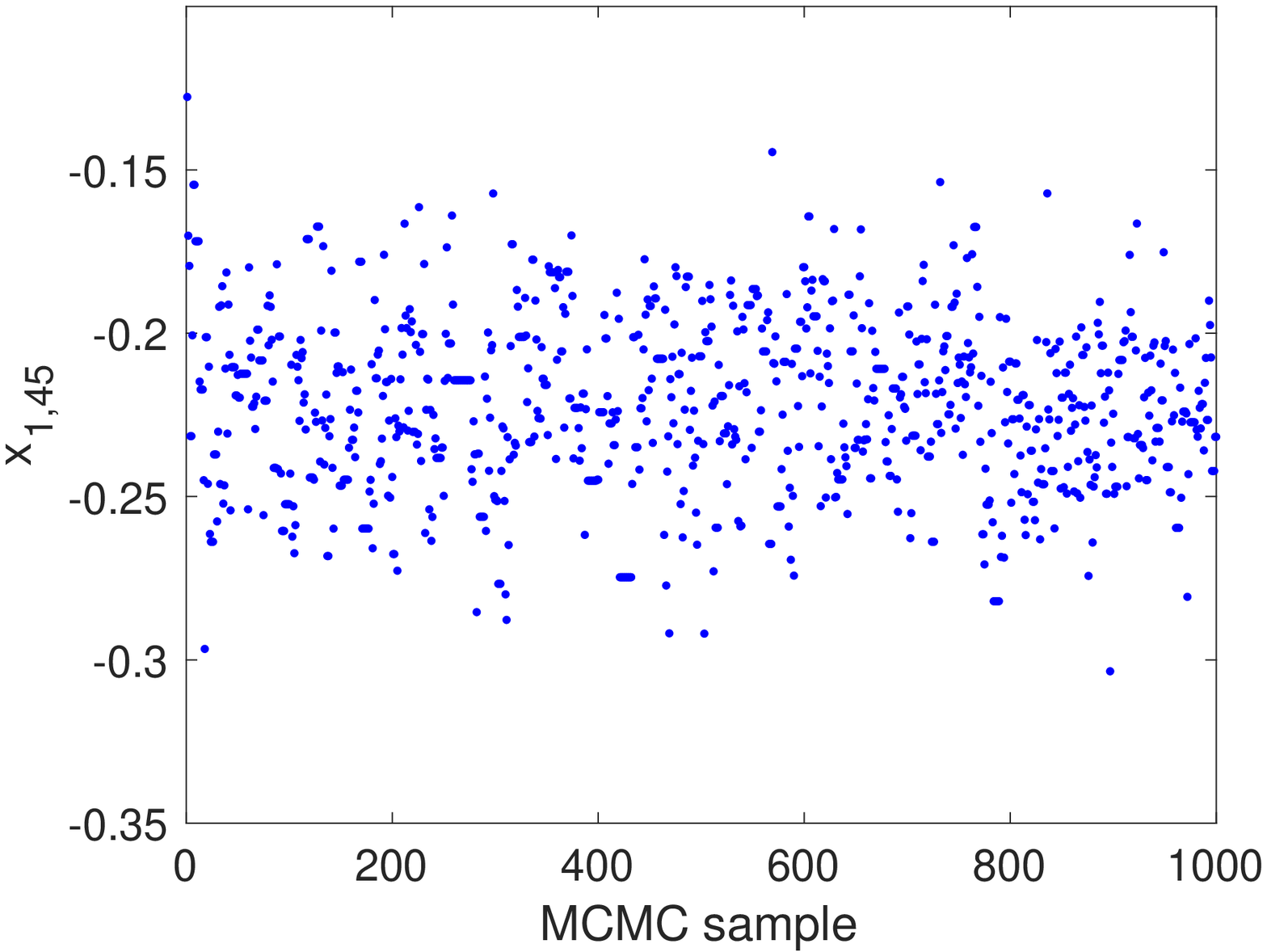}
\par\end{centering}
}
\par\end{centering}
\caption{Lorenz-96 model. Comparison of standard cSMC and replica cSMC.\label{fig:Lorenz-iterated-replica}}
\end{figure}

Figure \ref{fig:Lorenz-iterated-replica} shows the difference in
performance of the two samplers by trace plots of $x_{1,45}$ (true
value $-0.23$), from one of the runs, plotting the samples every
$30$th iteration. We can see that replica cSMC performs noticeably
better when compared to standard iterated cSMC.

\section{Conclusion}

We presented a novel sampler for latent sequences of a non-linear
state space model. Our proposed method leads to several
questions. The first is whether there are other ways to estimate the
predictive density that does not result in mixture weights with high
variance. Another question is to develop better guidelines on choosing
the number of replicas to use in a given scenario. It would also be
interesting to look at applications of replica cSMC in non time-series
examples. Finally, while the proposed method offers an approach for
sampling in models with multimodal state distributions, further improvement
is needed.

\clearpage
\bibliography{submission}

\begin{thebibliography}{13}
\providecommand{\natexlab}[1]{#1}
\providecommand{\url}[1]{\texttt{#1}}
\expandafter\ifx\csname urlstyle\endcsname\relax
  \providecommand{\doi}[1]{doi: #1}\else
  \providecommand{\doi}{doi: \begingroup \urlstyle{rm}\Url}\fi

\bibitem[Andrieu et~al.(2010)Andrieu, Doucet, and
  Holenstein]{Andrieu_Doucet_Holenstein_2010}
Andrieu, C., Doucet, A., and Holenstein, R.
\newblock Particle {M}arkov chain {M}onte {C}arlo (with discussion).
\newblock \emph{Journal of the Royal Statistical Society: Series B (Statistical
  Methodology)}, 72\penalty0 (4):\penalty0 269--342, 2010.

\bibitem[Briers et~al.(2010)Briers, Doucet, and Maskell]{Briers2010}
Briers, M., Doucet, A., and Maskell, S.
\newblock Smoothing algorithms for state-space models.
\newblock \emph{Annals of the Institute of Statistical Mathematics},
  62\penalty0 (1):\penalty0 61--89, 2010.

\bibitem[Chen et~al.(2013)Chen, Ming, and Liu]{Chen2013}
Chen, R., Ming, L., and Liu, J.~S.
\newblock Lookahead strategies for sequential {M}onte {C}arlo.
\newblock \emph{Statistical Science}, 28\penalty0 (1):\penalty0 69--94, 2013.

\bibitem[Finke et~al.(2016)Finke, Doucet, and Johansen]{Finke2016}
Finke, A., Doucet, A., and Johansen, A.
\newblock On embedded hidden {M}arkov models and particle {M}arkov chain
  {M}onte {C}arlo methods.
\newblock Technical report, arXiv:1610.08962, 2016.

\bibitem[Grothe et~al.(2016)Grothe, Kleppe, and
  Liesenfeld]{GrotheKleppeLiesenfeld}
Grothe, O., Kleppe, T., and Liesenfeld, R.
\newblock Bayesian analysis in non-linear non-{G}aussian state-space models
  using particle {G}ibbs.
\newblock Technical report, arXiv:1601.01125, 2016.

\bibitem[Guarniero et~al.(2017)Guarniero, Johansen, and
  Lee]{Guarniero_Lee_Johansen_2017}
Guarniero, P., Johansen, A.~M., and Lee, A.
\newblock The iterated auxiliary particle filter.
\newblock \emph{Journal of the American Statistical Association}, 112\penalty0
  (520):\penalty0 1636--1647, 2017.

\bibitem[Heng et~al.(2017)Heng, Bishop, Deligiannidis, and Doucet]{Heng2017}
Heng, J., Bishop, A., Deligiannidis, G., and Doucet, A.
\newblock Controlled sequential {M}onte {C}arlo.
\newblock Technical report, arXiv:1708.08396, 2017.

\bibitem[Leimkuhler et~al.(2018)Leimkuhler, Matthews, and
  Weare]{Leimkuhler2018}
Leimkuhler, B., Matthews, C., and Weare, J.
\newblock Ensemble preconditioning for {M}arkov chain {M}onte {C}arlo
  simulation.
\newblock \emph{Statistics and Computing}, 28:\penalty0 277--290, 2018.

\bibitem[Neal(2010)]{Neal2011}
Neal, R.
\newblock {MCMC} using ensembles of states for problems with fast and slow
  variables such as {G}aussian process regression.
\newblock Technical report, arXiv:1101.0387, 2010.

\bibitem[Ruiz \& Kappen(2017)Ruiz and Kappen]{Ruiz_Kappen_2017}
Ruiz, H.~C. and Kappen, H.~J.
\newblock Particle smoothing for hidden diffusion processes: adaptive path
  integral smoother.
\newblock \emph{IEEE Transactions on Signal Processing}, 65\penalty0
  (12):\penalty0 3191--3203, 2017.

\bibitem[Scharth \& Kohn(2016)Scharth and Kohn]{ScharthKohn2016}
Scharth, M. and Kohn, R.
\newblock Particle efficient importance sampling.
\newblock \emph{J. Econometrics}, 190\penalty0 (1):\penalty0 133--147, 2016.

\bibitem[Shestopaloff \& Neal(2018)Shestopaloff and Neal]{ShestopaloffNeal2018}
Shestopaloff, A. and Neal, R.
\newblock Sampling latent states for high-dimensional non-linear state space
  models with the embedded {HMM} method.
\newblock \emph{Bayesian Analysis}, 13\penalty0 (3):\penalty0 797--822, 2018.

\bibitem[Whiteley(2010)]{Whiteley2010}
Whiteley, N.
\newblock Discussion of particle {M}arkov chain {M}onte {C}arlo methods.
\newblock \emph{Journal of the Royal Statistical Society: Series B (Statistical
  Methodology)}, 72\penalty0 (4):\penalty0 306--307, 2010.

\end{thebibliography}
\bibliographystyle{icml2019}

\end{document}


\twocolumn[
\icmltitle{Supplementary Material for Replica Conditional Sequential Monte Carlo}



\icmlsetsymbol{equal}{*}

\begin{icmlauthorlist}
\icmlauthor{Alexander Y. Shestopaloff}{ed,turing}
\icmlauthor{Arnaud Doucet}{ox,turing}
\end{icmlauthorlist}

\icmlaffiliation{turing}{The Alan Turing Institute, London, UK}
\icmlaffiliation{ox}{Department of Statistics, University of Oxford, Oxford, UK}
\icmlaffiliation{ed}{School of Mathematics, University of Edinburgh, Edinburgh, UK}

\icmlcorrespondingauthor{Alexander Y. Shestopaloff}{ashestopaloff@turing.ac.uk}

\icmlkeywords{Markov Chain Monte Carlo, State Space Models}

\vskip 0.3in
]



\printAffiliationsAndNotice{}  

\section{Validity of Replica cSMC}

It is easy to see that the proposed update leaves $\bar{\pi}$ invariant.
Let $M_{x_{1:T}^{(-k)}}(x_{1:T}^{(k)'}|x_{1:T}^{(k)})$ be the cSMC
transition kernel used to update replica $x_{1:T}^{(k)}$, $k=1,\ldots,K$,
where $x_{1:T}^{(-k)}:=(x_{1:T}^{(1)'},\ldots,x_{1:T}^{(k-1)'},x_{1:T}^{(k+1)},\ldots,x_{1:T}^{(K)})$.
The replica update is a composition of the $M_{x_{1:T}^{(-k)}}$ so
we can write the replica cSMC transition kernel $M$ as a product,
$M(x_{1:T}^{(1:K)'}|x_{1:T}^{(1:K)})=\prod_{k=1}^{K}M_{x_{1:T}^{(-k)}}(x_{1:T}^{(k)'}|x_{1:T}^{(k)})$.

The replica cSMC transition kernel $M$ then leaves $\bar{\pi}$ invariant
since we have
\begin{align*}
& \int\bar{\pi}(x_{1:T}^{(1:K)})M(x_{1:T}^{(1:K)'}|x_{1:T}^{(1:K)})dx_{1:T}^{(1:K)} \\
 & =\int\prod_{k=1}^{K}p(x_{1:T}^{(k)}|y_{1:T})M_{x_{1:T}^{(-k)}}(x_{1:T}^{(k)'}|x_{1:T}^{(k)})dx_{1:T}^{(1:K)}\\
 & =\int\biggl[\int p(x_{1:T}^{(1)}|y_{1:T})M_{x_{1:T}^{(-1)}}(x_{1:T}^{(1)'}|x_{1:T}^{(1)})dx_{1:T}^{(1)}\biggr]\\
 & \times\prod_{k=2}^{K}p(x_{1:T}^{(k)}|y_{1:T})M_{x_{1:T}^{(-k)}}(x_{1:T}^{(k)'}|x_{1:T}^{(k)})dx_{1:T}^{(2:K)}\\
 & =p(x_{1:T}^{(1)'}|y_{1:T})\int\biggl[\int p(x_{1:T}^{(2)}|y_{1:T})M_{x_{1:T}^{(-2)}}(x_{1:T}^{(2)'}|x_{1:T}^{(2)})dx_{1:T}^{(2)}\biggr]\\
 & \times\prod_{k=3}^{K}p(x_{1:T}^{(k)}|y_{1:T})M_{x_{1:T}^{(-k)}}(x_{1:T}^{(k)'}|x_{1:T}^{(k)})dx_{1:T}^{(3:K)}\\
 & =p(x_{1:T}^{(1)'}|y_{1:T})p(x_{1:T}^{(2)'}|y_{1:T})\\
 & \times\int\prod_{k=3}^{K}p(x_{1:T}^{(k)}|y_{1:T})M_{x_{1:T}^{(-k)}}(x_{1:T}^{(k)'}|x_{1:T}^{(k)})dx_{1:T}^{(3:K)}\\
 & =\prod_{k=1}^{K}p(x_{1:T}^{(k)'}|y_{1:T})\quad(\textnormal{by induction)}\\
 & =\bar{\pi}(x_{1:T}^{(1:K)'}).
\end{align*}